\documentclass[11pt,a4paper]{article}
\usepackage{jcappub}

\usepackage{color}
\input{colordvi.tex}

\usepackage{amsmath}
\usepackage{graphicx}
\usepackage{float}
\usepackage{amsmath}

\usepackage{amsmath,amssymb}
\usepackage{graphicx}
\usepackage{subfigure}
\usepackage{verbatim}
\usepackage{makeidx}
\usepackage{amssymb}
\usepackage{ulem}

\newcommand{\Gyr}{\,{\rm Gyr}}
\newcommand{\Mpc}{\,{\rm Mpc}}

\title{Constraints on decaying dark matter from weak lensing and cluster counts}

\author[a]{Kari Enqvist,}
\author[b]{Seshadri Nadathur,}
\author[c]{Toyokazu Sekiguchi}
\author[d]{and Tomo Takahashi}
\affiliation[a]{University of Helsinki and Helsinki Institute of Physics, 
P.O. Box 64, FI-00014, Helsinki, Finland}
\affiliation[b]{Institute of Cosmology and Gravitation, 
University of Portsmouth, Burnaby Road, Portsmouth PO1 3FX, UK}
\affiliation[c]{Research Center for the Early Universe (RESCEU), 
Graduate School of Science, The University of Tokyo, 
Tokyo 113-0033, Japan}
\affiliation[d]{Department of Physics, Saga University, Saga 840-8502, Japan}

\emailAdd{kari.enqvist@helsinki.fi}
\emailAdd{seshadri.nadathur@port.ac.uk}
\emailAdd{sekiguti@resceu.s.u-tokyo.ac.jp}
\emailAdd{tomot@cc.saga-u.ac.jp}

\abstract{
We revisit a
cosmological constraint on dark matter decaying into dark radiation at late times.
In Enqvist et al. (2015), we mainly focused on the effects of decaying dark matter (DDM) 
on the cosmic microwave background (CMB) and nonlinear matter power spectrum.
Extending our previous analysis, here we use N-body simulation to investigate how DDM
affects the halo mass function. This allows us to incorporate the
cluster counts observed by the Sunyaev-Zel'dovich effect to study a bound
 on the lifetime 
of DDM. We also update the data of CMB and cosmic shear power spectrum with the Planck 
2015 results and KiDS450 observations, respectively.
From these cosmological observations, we obtain an lower bound on the lifetime $\Gamma^{-1}\ge 175\Gyr$ 
from the Planck2015 results (CMB+SZ cluster count) combined with the KiDS450 and the recent measurements of 
the baryon acoustic scale. 
}


\begin{document}

\begin{flushright}
RESCEU-8/19
\end{flushright}

\maketitle
\flushbottom

\section{Introduction} 
\label{sec:introduction}

Dark matter (DM) is one of the most important building blocks of the $\Lambda$ cold dark matter ($\Lambda$CDM) model, which is the standard paradigm of modern cosmology. 
DM makes up about 25 \% of the present Universe and DM particles should be stable over the age of the Universe.  However, 
this does not necessarily mean that they are perfectly stable.  In fact
decaying dark matter (DDM) can be realized in a broad class of particle physics models. Such decay would give a significant impact on 
various aspects of astrophysics and cosmology such as cosmic rays, the cosmic microwave background (CMB), large scale structure and so on. As a result, 
a lot of work has been devoted to investigate DDM models and their observational consequences. 
In particular, DDM models have attracted attention recently as it has been suggested that DDM can relax cosmological tensions
between the CMB and low-redshift observations, such as in the recovered values of $\sigma_8$ and the Hubble constant \cite{Berezhiani:2015yta,Enqvist:2015ara,Chudaykin:2016yfk,Poulin:2016nat,Bringmann:2018jpr,Pandey:2019plg,Vattis:2019efj}\footnote{For specific and motivated models beyond Standard model, 
we refer to {\it e.g.} \cite{Daido:2016tsj,Buch:2016jjp,Hamaguchi:2017ihw}.}.
 In \cite{Enqvist:2015ara}, we investigated this issue by using 
CMB data from the Planck 2013 data release \cite{Ade:2013kta} and weak lensing shear from CFHTLens \cite{Heymans:2013fya}, and showed that the tension in $\sigma_8$ between CMB and weak lensing survey can be 
alleviated to some extent by DDM.
 
In this paper, we extend the work of  \cite{Enqvist:2015ara} with the recent weak lensing data from KiDS450 \cite{Hildebrandt:2016iqg,Joudaki:2016kym} along with the Planck 2015 data \cite{Aghanim:2015xee}
and the Planck CMB lensing spectrum \cite{Ade:2015zua}. We also include other low-redshift observations from the Sunyaev-Zeldovich (SZ) cluster count from Planck \cite{Ade:2015fva} and baryon acoustic 
oscillation scales \cite{Beutler:2011hx,Ross:2014qpa,Anderson:2013zyy}. We assume that all DM  decays with the same decay rate $\Gamma$ 
and do not consider a mixed model ({\it i.e.}, a CDM + DDM model). 
As we argue in this paper, although KiDS450 and the SZ cluster count from Planck hint at a lower value of $\sigma_8$ compared to Planck, 
when we include multiple data from low redshift observations, the DDM model does not give a much better fit to the data, and is rather severely constrained. 
This is due to the fact that the different low-redshift observations are sensitive to different scales and redshifts and the DDM model cannot fit the data overall for a given decay rate $\Gamma$.\footnote{
Compared to Planck 2015, the use of the recent Planck 2018 data ~\cite{Aghanim:2018eyx}
(Planck TT,TE,EE+LowE+lensing) would make this tendency more noticeable as the later relase prefers a slightly lower $\sigma_8$ and is more
compatible with updated cosmic shear measurements~\cite{Abbott:2017wau,Hikage:2018qbn}.
However, the likelihood code for Planck 2018 has not yet been publicly released, so for this work we  use the 2015 likelihood and data release.} Therefore in this paper we aim to obtain a constraint on the decay rate of DDM by using the above mentioned data set rather than pursuing a possibility of 
resolving the tension of $\sigma_8$. The analysis is done by extending our previous work  \cite{Enqvist:2015ara}, but using a halo mass function 
calibrated from N-body simulation, which allows us to include the SZ cluster count data in our analysis. 
This is the new ingredient in the present paper. 
As will be shown in the following, the inclusion of the SZ cluster count provides a significant effect constraint on the dark matter decay rate.

Regarding the halo mass function in DDM model,  
let us comment on the differences between the one obtained in our present work and in previous studies. 
Ref.~\cite{Oguri:2003nn} studied the halo mass function in the same DDM model as ours,
based on an analytical argument. They argued that the abundance of cluster-sized halos is suppressed
in the DDM model and the deviation from standard CDM becomes prominent at later times, which agrees with 
our result. On the other hand, Refs.~\cite{Peter:2010au,Peter:2010jy}
also considered effects on halo mass function in DDM models, but where the decay products are
not massless. Refs.~\cite{Cheng:2015dga} studied the mass-concentration relation
as well as the mass function in DDM models with massive decay products.

Our paper is organized as follows. In the next section, we present the DDM model we consider in this paper.
We also briefly describe the methodology of our analysis.
In Section~\ref{sec:halo}, we discuss the effects of the DDM model 
on the halo mass function, based on which in Section~\ref{sec:constraints} we derive constraints on the 
DDM from recent cosmological observations.
We conclude in Section~\ref{sec:conclusion}. Appendices~\ref{app:fit} and \ref{app:c} respectively
describe our fitting formula for the DDM halo mass function, and the effects on the concentration of the haloes.

\section{Model and methodology} 
\label{sec:model}

Here we briefly describe the model we consider in this paper. Our methodology in this paper is the same as that in our previous paper~\cite{Enqvist:2015ara}, so we provide only a brief recap here, and
refer the readers to \cite{Enqvist:2015ara} for further details.

We investigate cosmological constraints on 
dark matter decaying into dark radiation (DR).
In the following, we assume that all dark matter particles decay,
and hence 
our decaying dark matter (DDM) model is characterized by only a 
single parameter, the decay time $\Gamma^{-1}$. We particularly focus on DDM with $\Gamma^{-1}$ 
larger than the age of the Universe. 

We assume a flat Universe consisting of 
photons, neutrinos, baryons, a cosmological constant~($\Lambda$), DDM and DR, which we call $\Lambda$DDM model hereafter.
Initially, dark radiation is assumed to be absent and is created only by the decay of DDM.
Thus the expansion history of the Universe is uniquely specified
once we specify $\Gamma$ and the abundance of baryons, DDM and the cosmological constant at some reference time. 
Since the energy densities of DDM and DR are not explicit functions of the scale factor $a$, 
it is more convenient to specify abundances of these constituents at the initial time ($a=0$)\footnote{
In practice, we start the calculation at $a \sim 10^{-7}$, well before the matter radiation equality. 
}
than at present ($a=1$).
Following Ref.~\cite{Enqvist:2015ara},
we adopt the parameters $\omega_i$ to specify the initial density of a constituent $i$, defined as:

\begin{equation}
\omega_i \equiv \left.\frac{\bar \rho_i (a) a^{3(1+\alpha_i)}}{\rho_{\rm crit}/h^2}\right|_{a=0}.
\end{equation}

Here, $\bar\rho_i$ and $\alpha_i$ are respectively the mean energy density 
and the equation of state of constituent $i$, and 
$\rho_{\rm crit}/h^2\equiv3\,(H_0/h)^2/8\pi G$
is the 
present critical density of the Universe, with $H_0=100h$\,km/sec/Mpc the Hubble parameter  
 and $G$ Newton's constant. Note that $\rho_{\rm crit}/h^2$ is a constant and does not depend on cosmology.
If the constituent $i$ is stable ({\it i.e.} photons, neutrinos, baryons or $\Lambda$),
$\omega_i$ coincides with the present density parameter $\Omega_i h^2$.
In addition, we define $h_\emptyset$, which represents the value of $h$
in the absence of dark matter decay. 

To investigate constraints from the CMB and weak lensing, 
we need to follow the evolution of cosmological perturbations in the model, 
which can be divided into linear and nonlinear regimes.

To study the evolution of linear perturbations in the $\Lambda$DDM model,
we modified the {\tt CAMB} Boltzmann code~\cite{Lewis:1999bs} to accommodate the effects of DDM.
The linear perturbation evolution in DDM models has already been studied by 
many authors \cite{Flores:1986jn,Takahashi:2003iu,Ichiki:2004vi,
Wang:2010ma,Aoyama:2011ba,Aoyama:2014tga,Audren:2014bca}.
We formulated a refined treatment of free streaming of the decay product, incorporating the approximations developed for neutrinos Ref.~\cite{Blas:2011rf}. 
Thus our study allows more accurate and fast computations of linear perturbation evolution in DDM.
This gives the initial condition for the N-body simulation as well as 
the angular power spectrum of the CMB.

We also investigate the nonlinear evolution of perturbations in DDM based on purpose-built N-body simulations.
Our N-body simulation incorporates two primary effects.
The first is the change to the background expansion which is caused by the transformation
of energy from DM to DR. The other effect is to allow DM to decay 
by making the mass of simulation particles time-dependent as
\begin{equation}
m(t)=m_i\{(1-r_{\rm dm})+r_{\rm dm} e^{-\Gamma t}\},
\end{equation}
where $m_i$ is the initial particle mass and 
$r_{\rm dm}$ is the fraction of mass density in total matter (dark matter + baryons).
We modified the publicly available {\tt Gadget-2} code~\cite{Springel:2000yr,Springel:2005mi} 
to incorporate these two primary effects.\footnote{
For a general relativistic treatment of N-body simulation we refer to~\cite{Dakin:2019dxu}.
So long as our analysis focuses on scales much smaller than horizon, general relativistic corrections are
subdominant.
}
Although our simulation omits the effects of perturbations in the DR produced by the decay, we have confirmed that 
this approximation is accurate enough for our analysis. 
To do this we checked the agreement of the power spectrum obtained from N-body simulation 
with the linear perturbation calculation from {\tt CAMB} at sub-horizon and but still linear scales, where both approaches should be valid.

When investigating constraints from weak lensing, we make use of the fitting function for the nonlinear matter power spectrum presented in \cite{Enqvist:2015ara}.

\section{Effects on the halo mass function}
\label{sec:halo}

In the DDM model, well before the decay time overdensities grow via gravitational instability 
as in the ordinary CDM model. 
At late times, as DDM decays  
the overdensities and the gravitational wells surrounding them begin
to fade away since the decay product, {\it i.e.} DR, is massless and 
leaks from overdense regions into underdense ones.
This moderates the gravitational instability and slows down structure 
formation. As mentioned in the previous section, 
the effects on the matter power spectrum, which is included in the fitting formula 
we obtained in our  previous paper~\cite{Enqvist:2015ara}, can be used to probe quasi-nonlinear scales. 
On the other hand, the formation of collapsed objects like dark matter haloes 
are also affected in the DDM model, which we evaluate in this section.

For this purpose, we performed the N-body simulations of collisionless particles
we had developed in our previous study~\cite{Enqvist:2015ara}.
We adopted three different box sizes, $L h_\emptyset/{\rm Mpc}=1250$, 500 and 200, and confirmed 
the convergence with resolution. 
For more details, we refer to Ref.~\cite{Enqvist:2015ara}.
Using the public {\tt AHF} halo-finding code \cite{Knollmann:2009pb}, haloes are identified based on
the spherical overdensity algorithm with $\Delta=500$.\footnote{
The overdensity is defined as fractional fluctuation in the energy density of DDM. Therefore, the
contribution of produced DR is omitted from the background energy density.}
Given the particle number $n$ contributing to a halo at time $t$, 
the mass of the halo is given by $M=m(t)n$. 

\begin{figure}
  \begin{center}
  \begin{tabular}{cc}
      \hspace{-5mm}\scalebox{.8}{\includegraphics{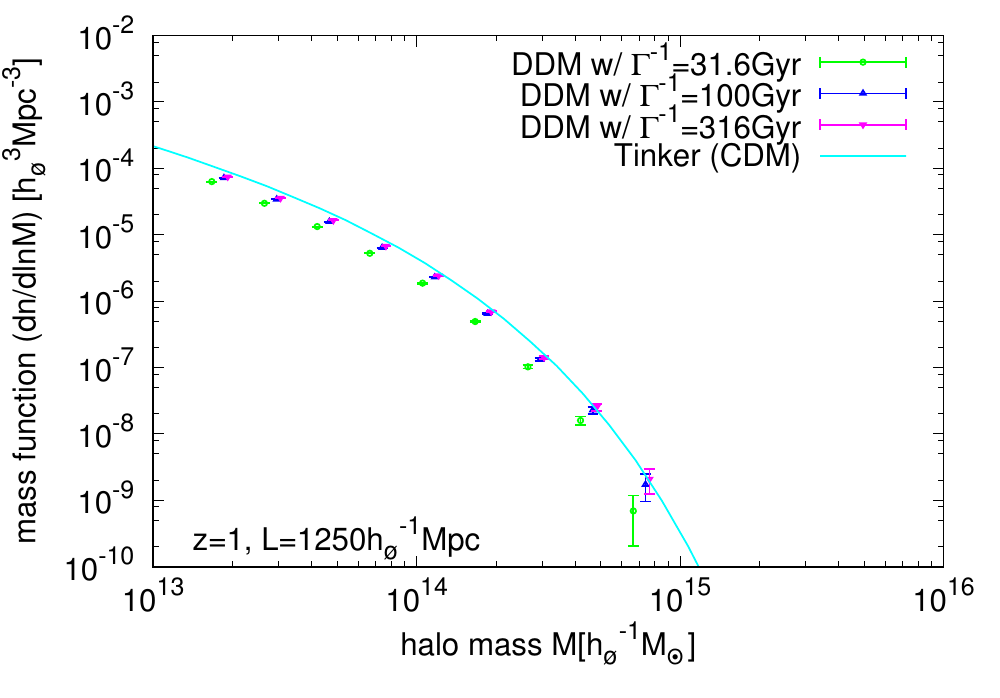}}
      & \hspace{-5mm}\scalebox{.8}{\includegraphics{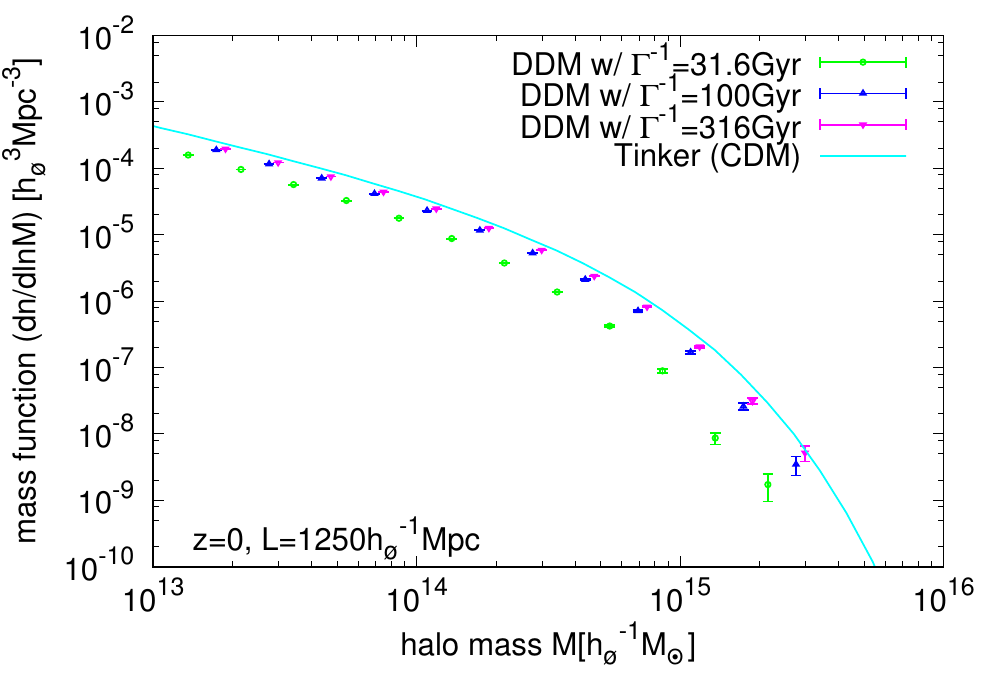}}\\
  \end{tabular}
  \end{center}
  \caption{\label{fig:comparison} Halo mass function in the DDM model at 
  redshifts $z=1$ (left) and $z=0$ (right) from simulation, shown for decay time $\Gamma^{-1}=31.6$ (green), $100$ (blue) and $316$Gyr (magenta). 
  For reference, we plot the Tinker mass function for the CDM model (cyan).
  The result is based on simulation with box size $L=1250h_\emptyset^{-1}\Mpc$
  and confirmed to be consistent with smaller box sizes $L=500$ and $200h_\emptyset^{-1}\Mpc$.
  }
\end{figure}

In figure~\ref{fig:comparison}, we plot the halo mass function in the DDM model. 
For reference, we also plot the Tinker mass function \cite{Tinker:2008ff}, which
well approximates the CDM prediction. The figure shows the suppression of 
the mass function relative the CDM model due to the dark matter decay. As one expects, the suppression becomes more
remarkable as the decay time $\Gamma^{-1}$ is decreased, and is more significant
at later times and larger halo masses.

From these simulation results, we developed a fitting formula for the 
halo mass function in the DDM model. Details of this fitting function are provided in Appendix~\ref{app:fit}.
We exploit the fitting formula over the range of parameters to cover an analysis for the SZ cluster count.

We  have also examined the effects on the halo inner profile. As summarized in Appendix~\ref{app:c},
deviations in the halo concentration parameter from CDM in the DDM model with $\Gamma^{-1}\ge 100$\,Gyr, which is our primary interest, 
are not significant in comparison with the variance among individual haloes. 
For the cluster-sized mass ({\it i.e.} $M\gtrsim 10^{14}M_\odot$), 
the impact of the DDM model on the concentration parameter is less prominent compared with the effects on 
the mass function we have seen above.
For the sake of clarity, in the analysis  we present in the next section, we omit the 
effects on the halo profile and only take into account effects on the mass function.

\section{Observational constraints}
\label{sec:constraints}
We here present updated constraints on the DDM model from recent cosmological observations.
We combine the CMB temperature and polarization power spectra \cite{Aghanim:2015xee} (hereafter CMB),
the CMB lensing spectrum (lensing) \cite{Ade:2015zua} and
the SZ cluster count \cite{Ade:2015fva} from the Planck 2015 results,
the cosmic shear power spectrum from KiDS450 \cite{Hildebrandt:2016iqg,Joudaki:2016kym}, 
and the compilation of the measurements of the baryon acoustic oscillation (BAO) scale 
\cite{Beutler:2011hx,Ross:2014qpa,Anderson:2013zyy}.
In order to obtain the posterior distribution of the cosmological parameters,
we use the publicly available {\tt CosmoMC} code~\cite{Lewis:2002ah} 
and modified it to incorporate the effects of the DDM on the cosmological observables.
Observables associated with the CMB, cosmic shear and BAO are computed 
as in the same manner as our previous study \cite{Enqvist:2015ara}. 
For example, we compute perturbation evolution of DDM in cosmological linear perturbation theory, 
which yields CMB angular power spectra. Cosmic shear power spectrum is computed using
the nonlinear matter power spectrum which we have established based on results of N-body simulation.
BAO is computed only by taking into account DDM effect on the background expansion.
In addition, when we use the SZ cluster count, we take into account the halo mass function in the DDM model given in Appendix \ref{app:fit}.

As mentioned in Section~\ref{sec:model}, we assume a flat power-law $\Lambda$DDM model, with dark matter 
assumed to be 100\% DDM. The primary cosmological parameters we vary are
$(\omega_b$, $\omega_{\rm ddm}$, $\tau_{\rm reion}$, $\theta_{s}$, $\log(10^{10}A_s)$, $n_s$, and $\Gamma)$,
where $\omega_b$ and  $\omega_{\rm ddm}$ are respectively the density parameters of 
baryon and DDM, $\tau_{\rm reion}$ is the reionization optical depth, 
$\theta_s$ is the angular size of sound horizon at last scattering, 
and $A_s$ and $n_s$ are respectively the amplitude and spectral index of 
the power spectrum of the primordial curvature perturbations.

\begin{figure}
  \begin{center}
      \hspace{-5mm}\scalebox{.75}{\includegraphics{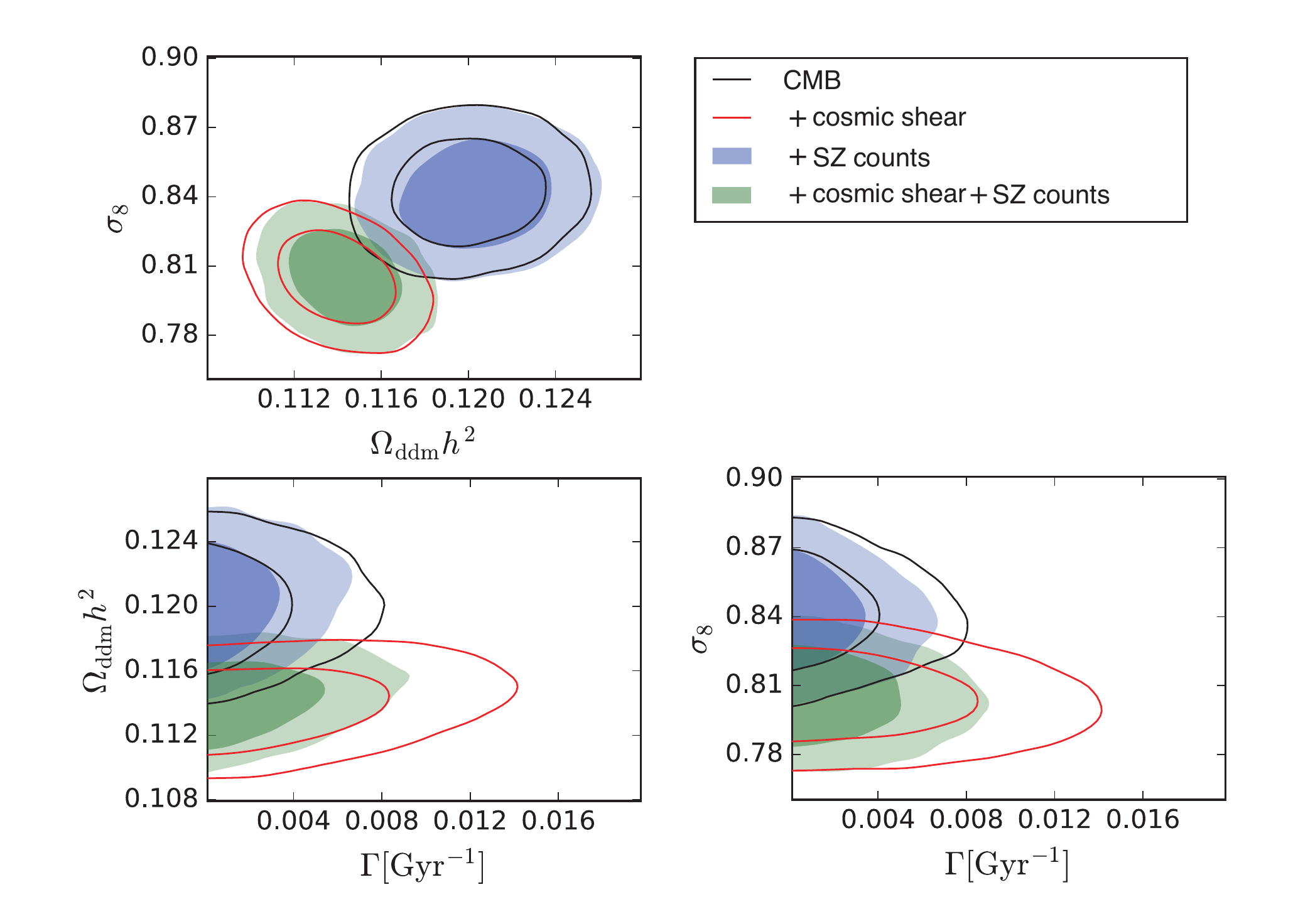}}
  \end{center}
  \caption{\label{fig:2d1} Two dimensional 68\% and 95\% confidence limit contours on $\Omega_{\rm ddm}$, $\sigma_8$ and $\Gamma$ from the different data combinations with CMB alone being the baseline.
  }
\end{figure}

\begin{figure}
  \begin{center}
      \hspace{-5mm}\scalebox{.75}{\includegraphics{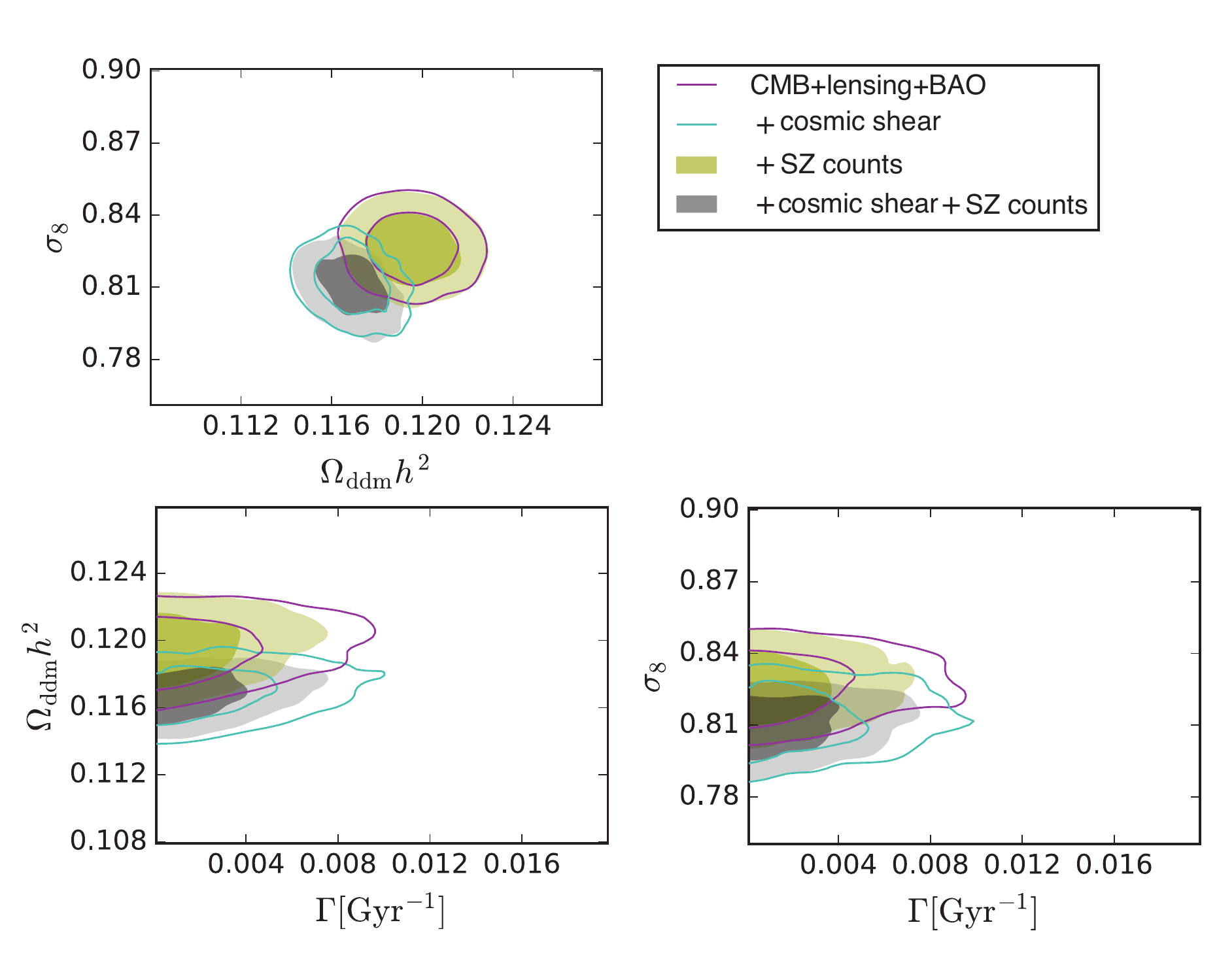}}
  \end{center}
  \caption{\label{fig:2d2} Same figure as in \ref{fig:2d1} but with CMB, lensing and BAO as baseline.
  }
\end{figure}

In Table~\ref{tab:limits} we summarize the constraints on
the DDM decay rate $\Gamma$. 
In figures~\ref{fig:2d1} and \ref{fig:2d2} we plot the 2D constraints projected onto the parameter space of $\omega_{\rm ddm} (= \Omega_{\rm ddm}h^2)$, $\sigma_8$ and $\Gamma$.
First of all, we found that the all the combination of the cosmological observations are consistent with
vanishing decay rate $\Gamma=0$. 
The loosest constraint, $\Gamma<1.1\times 10^{-2}\Gyr^{-1}$, comes from the combination of CMB with cosmic shear.
The preference for nonzero $\Gamma$ from the cosmic shear was also seen in our previous analysis~\cite{Enqvist:2015ara}
where we used the results from the CFHTLenS survey~\cite{Heymans:2013fya}. We confirmed the persistence of the preference 
in the KiDS450 data. On the other hand, the tightest constraint, $\Gamma<4.7\times 10^{-3}\Gyr^{-1}$, is obtained from the 
combination of CMB with SZ cluster count. Current SZ cluster count data slightly improves the bound on $\Gamma$ from CMB alone.

\begin{table}
    \begin{center}
    \begin{tabular}{lc}
    \hline
    \hline 
    & $\Gamma~[10^{-3}\Gyr^{-1}]$ \\
    CMB & $<6.3$\\
    \quad+lensing & $<7.0$\\
    \quad+cosmic shear & $<11$\\
    \quad+SZ clusters & $<4.9$\\
    \quad+cosmic shear+SZ clusters& $<6.8$\\
    CMB+BAO+lensing & $<7.3$\\
    \quad+cosmic shear & $<7.5$\\
    \quad+SZ clusters & $<5.5$\\
    \quad+cosmic shear+SZ clusters& $<5.7$\\
    \hline
    \hline
    \end{tabular}
    \caption{\label{tab:limits}  
    Constraints on $\Gamma$ from different cosmological datasets. 
    }
  \end{center}
\end{table}

In closing this section, let us discuss
implications for the tension reported in measured $H_0$ between
CMB and low-$z$ observations. 
It has been argued that DDM can solve the tension by previous studies \cite{Bringmann:2018jpr,Pandey:2019plg,Vattis:2019efj}.
The basic idea is as follows. Provided fixed expansion history of the flat Universe before $\Lambda$ domination, 
the expansion rate at late times increases if $h$ is increased or $\Gamma$ is decreased.
Therefore, larger $h$, which is indicated by local measurements can be in principle allowed by CMB with a nonzero $\Gamma$. 
Nonetheless, we found that allowed value of $h$ changes no more than one percent between CDM and DDM
since $\Gamma$ is tightly constrained by for example, late-time Sachs-Wolfe effect.
To significantly mitigate the tension in $H_0$, we need to generalized our DDM model furthermore
(for instance, massive decay product \cite{Berezhiani:2015yta,Chudaykin:2016yfk,Vattis:2019efj}, or mixture of CDM and DDM
\cite{Poulin:2016nat}).

\section{Conclusion}\label{sec:conclusion}
We have investigated the nonlinear structure formation in the DDM model, in particular focussing on the 
halo abundance. For this purpose, extending
our previous study~\cite{Enqvist:2015ara}, 
we  performed N-body simulations in the DDM model. We have shown that 
DDM suppresses the halo abundance in these simulations. This suppression is predominantly due to
the mass loss of the formed clusters originating from
the decay of DDM, while the relaxation
of the gravitational instability caused by the DM decay also contributes.
Adopting the fitting function for the halo abundance based on the simulation, 
we derived cosmological constraints on the DDM from the Planck 2015 SZ cluster count combined with
the Planck 2015 CMB power spectrum and the KiDS450 cosmic shear power spectrum.
We have found the cosmological observations are consistent with CDM and obtained 
a lower bound on the lifetime of DM as $\Gamma^{-1}\ge 175\Gyr$ from the combination of all the data above.

We note that our simulation is based on the collisionless N-body simulations and hence baryonic effects
are not taken into account. We expect that baryonic effects will not affect the mass function
of cluster-sized massive haloes very significantly. Moreover, since the baryonic effect in general decreases 
the mass of haloes and hence further suppresses the mass function for given mass~\cite{Cusworth:2013vtw,Bocquet:2015pva}, 
its effect is more or less degenerate with the decay rate of DDM. Therefore, one can regard our lower bound 
on $\Gamma^{-1}$ as conservative in respect to the baryonic effect.

\acknowledgments 
This research was supported by JSPS KAKENHI Grant Numbers 15K05084 (TT), 17H01131 (TT),  
15H02082 (TS), 18H04339 (TS), 18K03640 (TS),  MEXT KAKENHI Grant Number 15H05888 (TT)
and  UK Space Agency grant ST/N00180X/1.
We thank the CSC - IT Center for Science (Finland) for computational resources.

\appendix

\section{Fitting formula for halo mass function}
\label{app:fit}
We here present the fitting formula for the halo mass function in DDM model.
Given the DDM decay rate $\Gamma$, 
the suppression in the mass function from the CDM case ({\it i.e.} $\Gamma=0$)
at halo {\it initial} mass $M_i$ and redshift $z$ can be approximated by
\begin{equation}
\frac{(dn/dM_i)_{\rm DDM}}{(dn/dM_i)_{\rm CDM}}-1=a(\Gamma,z)\left[1+\left(\frac{M_i}{b(\Gamma,z)
10^{15}h_\emptyset M_\odot}\right)^{c(\Gamma,z)}\right]^{-1},
\end{equation}
where $a$, $b$, $c$ are functions of $\Gamma$ and $z$ given as
\begin{eqnarray}
a(\Gamma,z)&=&\exp\left[a_1\left(\frac \Gamma{\Gyr^{-1}}\right)+\frac{a_2}{1+z}\right], \\
b(\Gamma,z)&=&b_0 \left(\frac \Gamma{\Gyr^{-1}}\right)^{b_1}(1+z)^{b_2}, \\
c(\Gamma,z)&=&c_0 \left(\frac \Gamma{\Gyr^{-1}}\right)^{c_1}(1+z)^{c_2},
\end{eqnarray}
with 
\begin{eqnarray}
&a_1=-5.72,\quad a_2=7.56\times10^{-3}, \notag \\
&b_0=4.28\times10^{-4},\quad b_1=-2.34,\quad b_2= 0.567,\\
&c_0=1.16,\quad c_1=0.196,\quad c_2=-9.24\times 10^{-2}. \notag
\end{eqnarray}
We note that our fitting formula is calibrated with
the best-fit parameters of the Planck 2015 TT+TE+EE results.

\begin{figure}
  \begin{center}
  \begin{tabular}{cc}
      \hspace{-5mm}\scalebox{.7}{\includegraphics{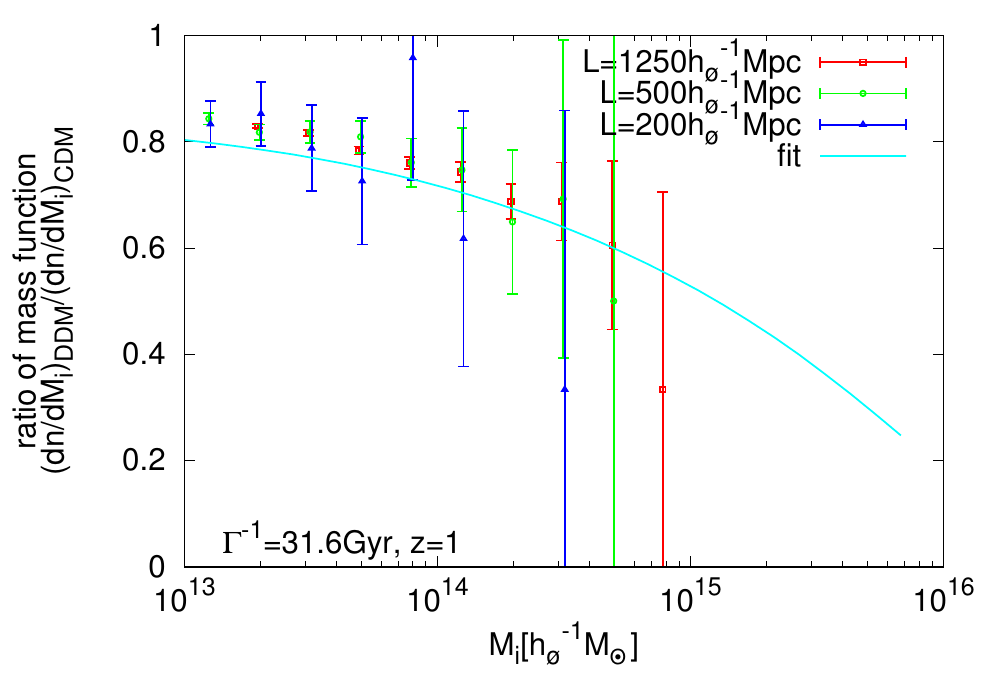}}
      & \hspace{-5mm}\scalebox{.7}{\includegraphics{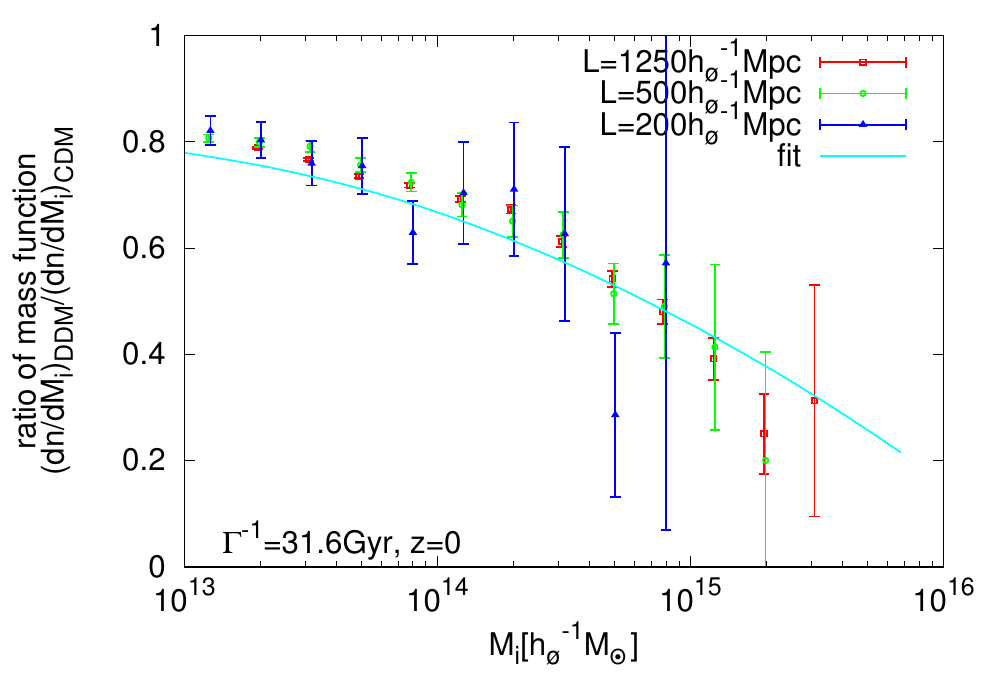}}\\
      \hspace{-5mm}\scalebox{.7}{\includegraphics{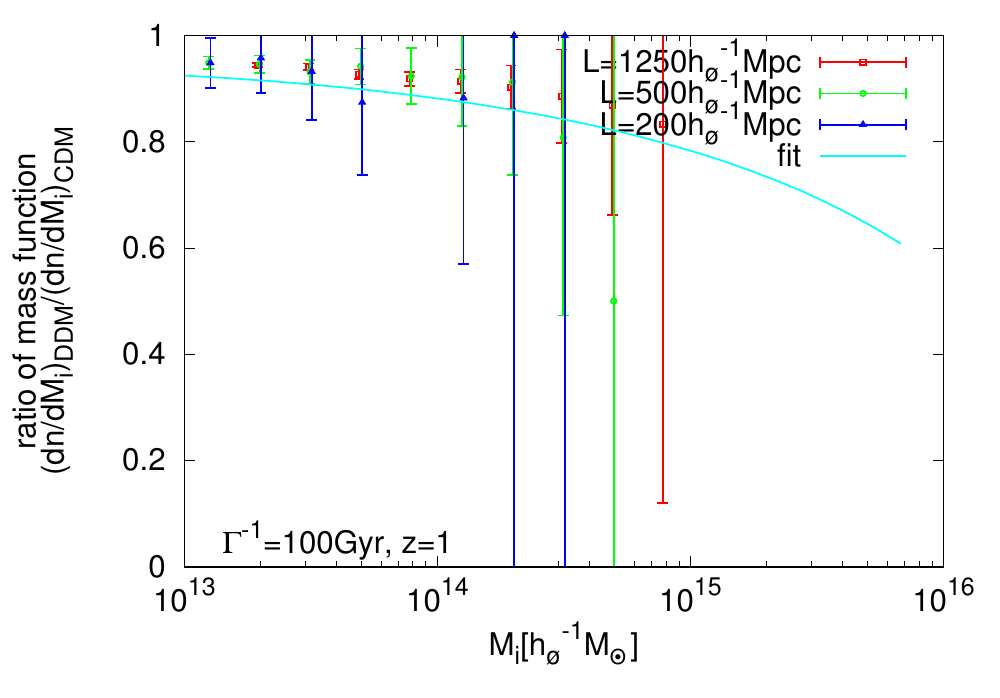}}
      & \hspace{-5mm}\scalebox{.7}{\includegraphics{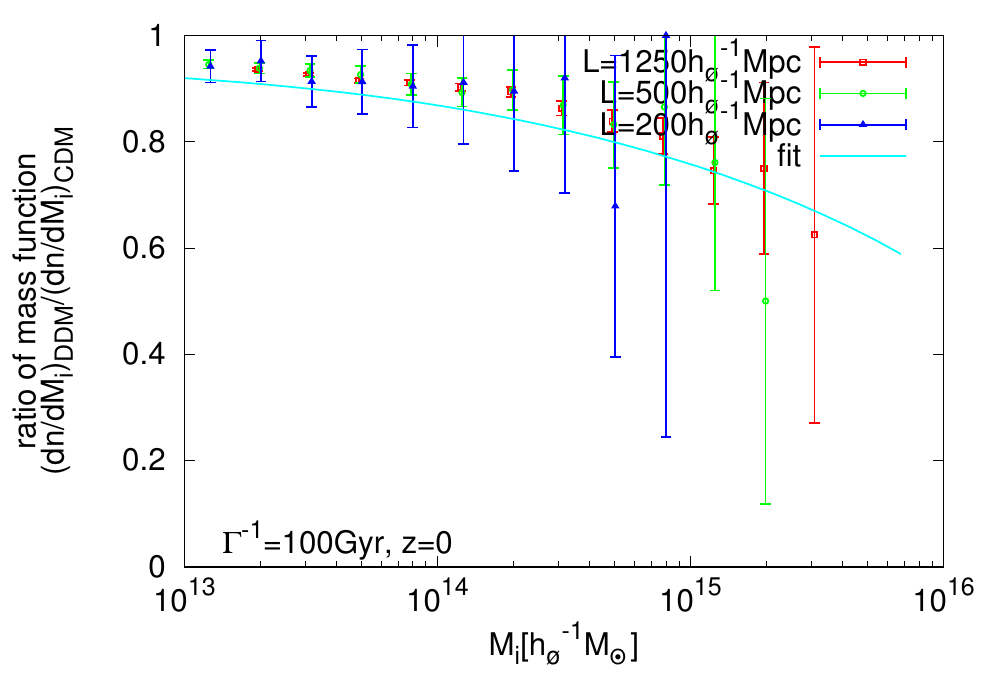}}\\
  \end{tabular}
  \end{center}
  \caption{
Ratio of the mass functions $ \left. \left( \frac{dn}{dM_i} \right) \right|_{\rm DDM}$ to $ \left. \left( \frac{dn}{dM_i} \right) \right|_{\rm CDM}$ for the cases  of
  $\Gamma^{-1}=$31.6~(top), $100\Gyr$~(bottom) for $z=1$ (left) and $z=0$ (right). 
  Here we show our fitting formula (cyan) in comparison with
  the results of our N-body simulations with box sizes $L h_\emptyset/{\rm Mpc}=1250$ (red),
  500 (green), 200 (blue). 
  }
  \label{fig:fitting}
\end{figure}

Our fitting formula can reproduce the suppression $\epsilon$ as function of $M_i$
with accuracy of $\sim20$\% for $10^{14}\le M_i/(h_\emptyset M_\odot)\le10^{15}$, 
$0\le z\le1$ for $\Gamma^{-1}=31\Gyr$ as is shown in the figure~\ref{fig:fitting}. 
For larger $\Gamma^{-1}$, the suppression in the mass function becomes 
less prominent compared to the statistical error, which makes it harder to 
assess the accuracy of our fitting formula in terms of the suppression factor $\epsilon$. 
Still, our fitting formula shows reasonable agreement with the simulation results.

Moreover, as shown in figure~\ref{fig:comparison}, 
the mass function in terms of actual halo mass $M=M_i\{(1-r_{dm})+r_{dm} e^{-\Gamma t}\}$ 
exhibits more prominent suppression from the CDM model than in the initial halo mass $M_i$.
This is because halo mass function is the decreasing function of $M_i$, and 
$M_i$ in the DDM model should necessarily be larger than that in the CDM model. 
Even if the abundance of halos with $M_i$ were the same in these two models,
given a fixed actual halo mass $M$, the mass function of $M$ in DDM would be suppressed 
compared to CDM. In reality, the halo mass function of $M_i$ in DDM model is suppressed relative to 
the one in the CDM model as shown in figure \ref{fig:fitting}.

%

\section{Effects on concentration}
\label{app:c}

\begin{figure}
  \begin{center}
  \begin{tabular}{cc}
      \hspace{-5mm}\scalebox{.5}{\includegraphics{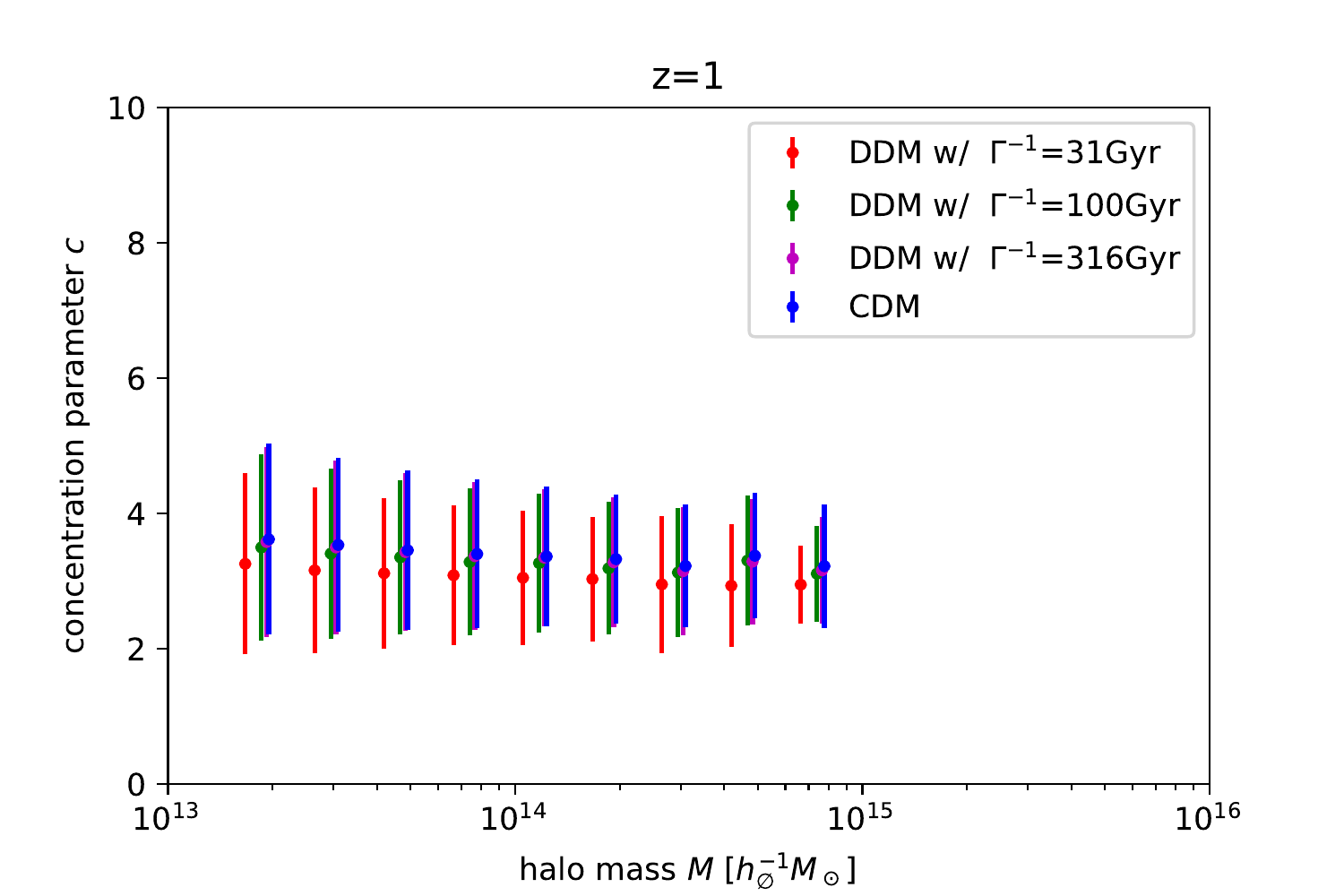}}
      & \hspace{-5mm}\scalebox{.5}{\includegraphics{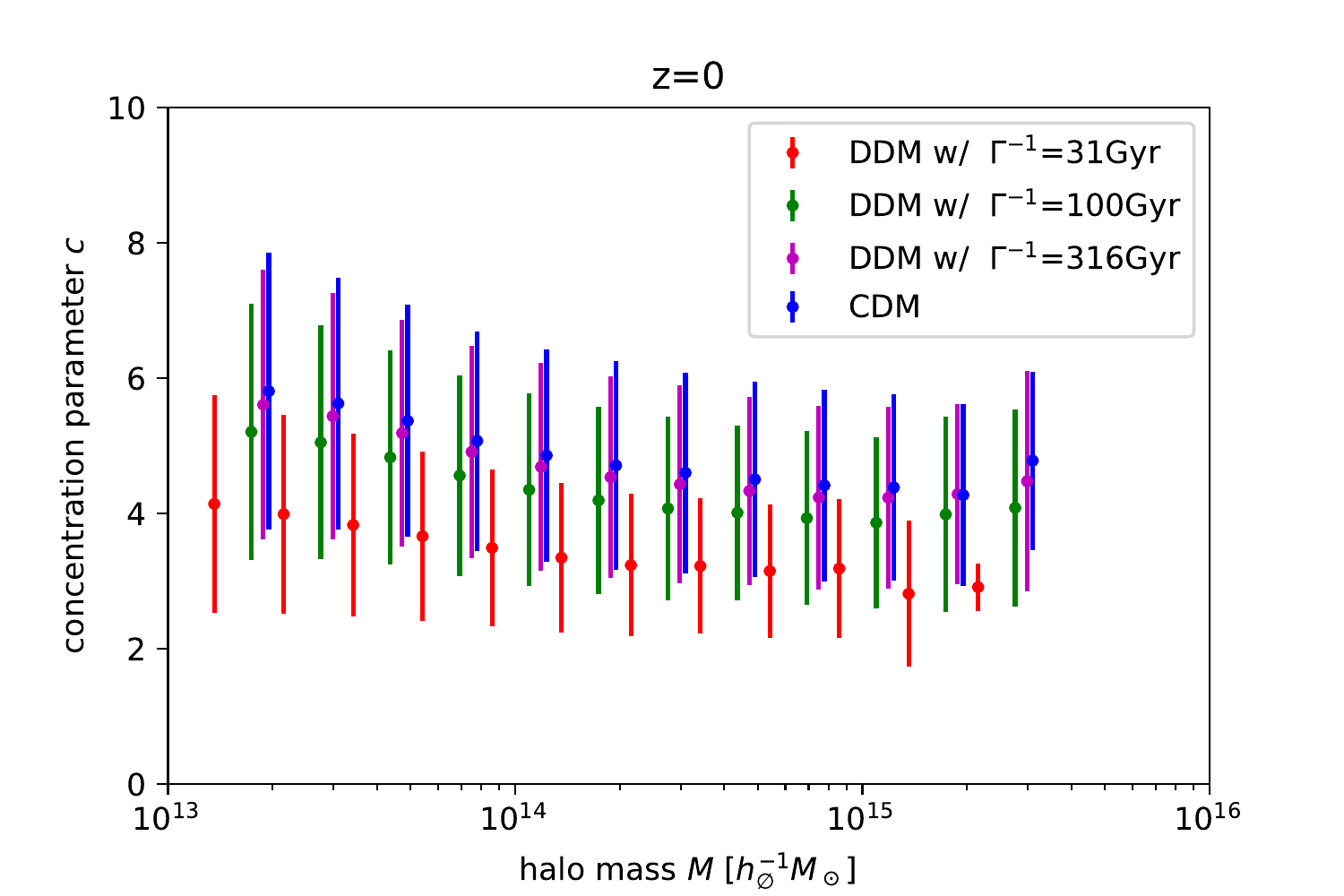}}
  \end{tabular}
  \end{center}
  \caption{Effects of the DDM on the halo concentration at redshifts $z=1$ (left) and 0 (right).
  Shown are the NFW concentration parameter $c$ for the DDM models with $\Gamma^{-1}=31$ (red),
  $100$ (green), $316$~Gyr (magenta), as well as for the CDM one (blue). Here we adopted the results 
  of our N-body simulations with a box size $L h_\emptyset/{\rm Mpc}=1250$. 
  }
  \label{fig:c}
\end{figure}

Figure \ref{fig:c} shows the concentration parameter $c$ as function of the halo mass.
We have fitted the density profile of halo assuming the NFW halo profile~\cite{Navarro:1995iw}:
\begin{equation}
\rho(r) = \frac{\rho_0}{c\frac{r}{R_{\rm vir}}\left(1+c\frac{r}{R_{\rm vir}}\right)^2}
\end{equation}
where $R_{\rm vir}$ is the halo virial radius and $\rho_0$ is the density at reference radius.
For $\Gamma^{-1}\ge 100$\,Gyr, we conclude the concentration parameter is suppressed from CDM
by about 10\% percent at $z=0$. For higher redshifts, the extent of the suppression becomes less. 
As Fig.~\ref{fig:comparison} shows that the halo mass function is suppressed by a factor of unity 
for cluster-sized halo mass ($M\gtrsim 10^{14}M_\odot$),  
the primary effects of DDM manifests in the halo mass function rather than the concentration.

\end{document}